\documentclass[preprint,11pt,pre,floats,aps,amsmath,amssymb]{revtex4}

\usepackage{graphicx}
\usepackage{bm}

\begin{document}

\title{Diffraction tomography with Fourier ptychography}


\author{R. Horstmeyer$^{*}$ and C. Yang}
\affiliation{Department of Electrical Engineering, California Institute of Technology, Pasadena, CA, 91125, USA}

\author{$^{*}$Corresponding author: roarke@caltech.edu   \vspace{10 mm}} 


\begin{abstract}
This article presents a method to perform diffraction tomography in a standard microscope that includes an LED array for illumination. After acquiring a sequence of intensity-only images of a thick sample, a ptychography-based reconstruction algorithm solves for its unknown complex index of refraction across three dimensions.  The experimental microscope demonstrates a spatial resolution of 0.39 $\mu$m and an axial resolution of 3.7 $\mu$m at the Nyquist-Shannon sampling limit (0.54 $\mu$m and 5.0 $\mu$m at the Sparrow limit, respectively), across a total imaging volume of 2.2 mm$\times$2.2 mm$\times$110 $\mu$m. Unlike competing methods, the 3D tomograms presented in this article are continuous, quantitative, and formed without the need for interferometry or any moving parts. Wide field-of-view reconstructions of thick biological specimens demonstrate potential applications in pathology and developmental biology.
\end{abstract}

\maketitle

\section{Introduction}
\label{sec:intro}

It is challenging to image thick samples with a standard microscope. High-resolution objective lenses offer a shallow depth-of-field, which require one to axially scan through the sample to visualize three-dimensional shape. Unfortunately, refocusing does not remove light from areas above and below the plane of interest. This longstanding problem has inspired a number of solutions, the most widespread being confocal designs, two-photon excitation methods, light sheet microscopy, and optical coherence tomography. These methods ``gate out" light from sample areas away from the point of interest, and offer excellent signal enhancement, especially for thick, fluorescent samples~\cite{Ntziachristos10}.

Such gating techniques also encounter several problems. First, they typically must scan out each image, which might require physical movement and can be time consuming. Second, the available signal (i.e., the number of ballistic photons) decreases exponentially with depth. To overcome this limit, one must use a high NA lens, which provides a proportionally smaller image field-of-view (FOV). Finally, little light is backscattered when imaging non-fluorescent samples that are primarily transparent, such as commonly seen in embryology, in model organisms such as zebrafish, and after the application of recent tissue-clearing~\cite{Chung13} and expansion~\cite{Chen15} techniques. 

Instead of capturing just the ballistic photons emerging from the sample, one might instead image the entire optical field, including the scattered components. This avoids point scanning, and allows one to record a very wide image FOV in a single snapshot. Several techniques have been proposed to enable depth selectivity after full-field capture. First, one might perform optical sectioning by capturing a focal stack, and then attempting digital deconvolution~\cite{Agard84}. A second related example is light-field imaging~\cite{Levoy09, Broxton13}. Point-spread function engineering is a third possibility~\cite{Pavani09}, but this typically requires a sparse sample. All three of these methods primarily operate with incoherent light, e.g. from fluorescent samples. They are thus not ideal tools for obtaining the complex refractive index distribution of a primarily transparent medium.
 
To do so, it is useful to use coherent illumination. For example, the amplitude and phase of a digital hologram may be computationally propagated to different depths within a thick sample, much like refocusing a microscope. However, the field at out-of-focus planes still influences the final result. Several techniques have improved upon depth selectivity with quasi-coherent illumination, based upon the acquisition of multiple images~\cite{Streibl85,Dubois02,Adie12, Kang15, Matthews14}. 

A very useful framework to summarize how coherent light scatters through thick samples is diffraction tomography (DT)~\cite{Wolf69}. This framework connects the optical fields that diffract from a sample, under arbitrary illumination, to its 3D complex refractive index. In a typical DT experiment, one illuminates a sample of interest with a series of tilted plane waves and measures the resulting complex diffraction patterns in the far field. These measurements may then be combined with a suitable algorithm into a tomographic reconstruction. As a synthetic aperture technique, DT comes with the additional benefit of improving the limited resolution of an imaging element beyond its traditional diffraction-limit cutoff~\cite{Lauer02}. Thus, it appears like a well-suited method to study thick, transparent samples at high resolution. 

However, as a technique that models both the amplitude and phase of a coherent field, nearly all prior implementations of DT required a reference beam and holographic measurement, or some sort of phase-stable interference (including SLM coding strategies, e.g. as in \cite{Kim14}). Reference fields require sub-micrometer stability in terms of both motion and phase drift, which has thus far limited DT to well-controlled, customized setups. While several prior works have considered solving the DT problem from intensity-only measurements from a theoretical perspective~\cite{Maleki93,Stamnes95,Gbur02,Anastasio05,Huang07}, none have implemented a DT system within a standard microscope, or connected their reconstruction attempts to ptychography. 

Here, we perform DT based upon intensity images captured under variable LED illumination from an array source. Our technique, termed Fourier ptychographic tomography (FPT), captures a sequence of images while changing the light pattern displayed on the LED array. Then, it combines these images using a phase retrieval-based ptychographic reconstruction algorithm, which computationally (as opposed to physically) rejects light from all areas above and below each plane of interest. FPT also improves the lateral image resolution beyond the standard cutoff of the objective lens used for imaging. The end result is a quantitatively accurate three-dimensional map of the complex index of refraction of a volumetric sample, obtained directly from a sequence of standard microscope images. 

\begin{figure}[]
\centering
\includegraphics[width=.9\columnwidth]{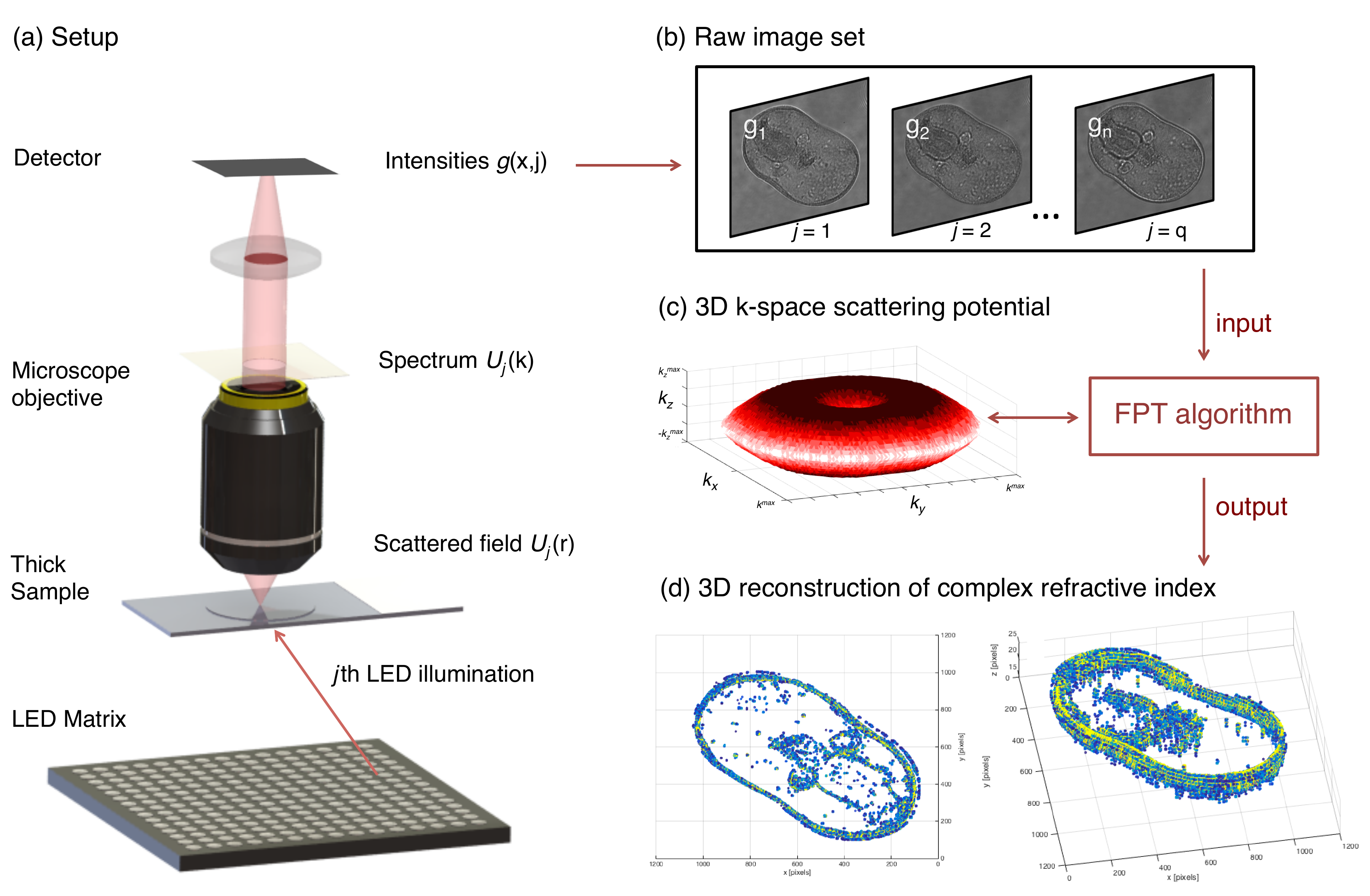}
\caption[2D]{Setup for Fourier ptychographic tomography (FPT). (a) A labeled diagram of the FPT microscope, with optical fields of interest labeled. (b) Multiple images are acquired under varied LED illumination. (c) A ptychography-inspired algorithm combines these images within a 3D k-space representation of the complex sample of interest. (d) FPT outputs a 3D tomographic map of the complex index of refraction of the sample. Included images are experimental measurements from a starfish embryo.}
\vspace{-0.2cm}
\label{teaser}
\end{figure}

\section{Related Work}
\label{sec:related}

The theoretical foundations of DT were first developed by Wolf~\cite{Wolf69}. A number of implementations based upon holography have followed. An early demonstration by Lauer is a good example~\cite{Lauer02}. Prior methods have also implemented tomography within a microscope-like setup, but required the addition of a phase-stable reference beam. Early results applied the projection approximation, which models light as a ray~\cite{Choi07}. Subsequent work has taken the effects of diffraction into account~\cite{Debailleul08,Debailleul09, Sung09}.

Instead of relying upon holography, this work measures intensity images and computationally recovers the missing phase of each field. As mentioned above, a few prior works consider the reconstruction problem from detected optical intensities, but must either move the focal plane between measurements, or must assume a sample support constraint. They do not attempt ptychographic phase retrieval. One related strategy worth mentioning is lifting-based phase retrieval for tomography~\cite{Chai11}. The connection between the first Born approximation and phase retrieval has also been explored within the context of volume hologram design~\cite{Gerke10}. 

Related efforts to reconstruct volumetric samples from wide-field intensity-only measurements outside of the realm of DT include lensless on-chip setups~\cite{Isikman11, Zuo15}, lensless techniques that assume an appropriate linearization~\cite{Gureyev06}, and methods relying upon effects like defocus (e.g., the transport of intensity equation~\cite{Bronnikov02}) or spectral variations~\cite{Kim2014a}. None of these techniques fit within a standard microscope setup, nor offer the ability to simultaneously improve spatial resolution. 

Based upon a standard microscope, Fourier ptychography (FP)~\cite{Zheng13} can simultaneously improve image resolution and measure quantitative phase~\cite{Ou13}. However, it is restricted to thin samples. FPT effectively extends prior developments of FP into the third dimension. Two recent works also examined the problem of 3D imaging from intensities in a standard microscope~\cite{Tian15, Li15}. These two examples adopted their 3D reconstruction technique from the related field of 3D ptychography~\cite{Maiden12, Godden14}, where the sample under examination is split up into a specified number of infinitesimally thin slices, and the beam propagation method (i.e., multi-slice approximation) is applied~\cite{Roey81}. Unlike the multi-slice approach, which works well with distinctly separated absorbing layers, FPT is best suited for continuous, primarily transparent samples. A number of related methods to perform 3D X-ray ptychography have also been proposed~\cite{Dierolf10,Peterson12,Shimomura15}. However, none seem to directly modify DT under the first Born or Rytov approximation, to the best of our knowledge. A popular technique appears to use standard 2D ptychographic solvers to determine the complex field for individual projections of a slowly rotated sample, which are subsequently combined using conventional DT techniques, as shown with both crystallographic~\cite{Hruszkewycz11} and unordered specimens~\cite{Putkunz10}. 

Here, we first outline a solid foundation for the application of ptychographic phase retrieval to DT. Unlike approaching the problem from a projection-based or multi-slice perspective, the framework of DT (under the first Born approximation) follows directly from the scalar wave equation. It offers a clear picture of achievable resolution in 3D, spells out sampling and data redundancy requirements for an accurate reconstruction, and presents a clear path forward for future extensions to account for multiple scattering~\cite{Wang89}. Furthermore, our method does not require the arbitrary assignment of the number slices in the 3D volume, or their location, or for us to select a particular order in which to address each slice as iterations proceed. Instead, it simply inserts the measured data into its appropriate location in 3D Fourier space and ensures phase consistency between each measurement, given a sufficient amount of data redundancy (just like ptychography). From the initial starting point of solving for the first term in the Born expansion, we aim this approach as a general framework to eventually solve the challenging problem of forming tomographic maps of volumetric samples, at sub-micrometer resolution, in the presence of significant scattering.  

\begin{figure}[]
\centering
\includegraphics[width=.75\columnwidth]{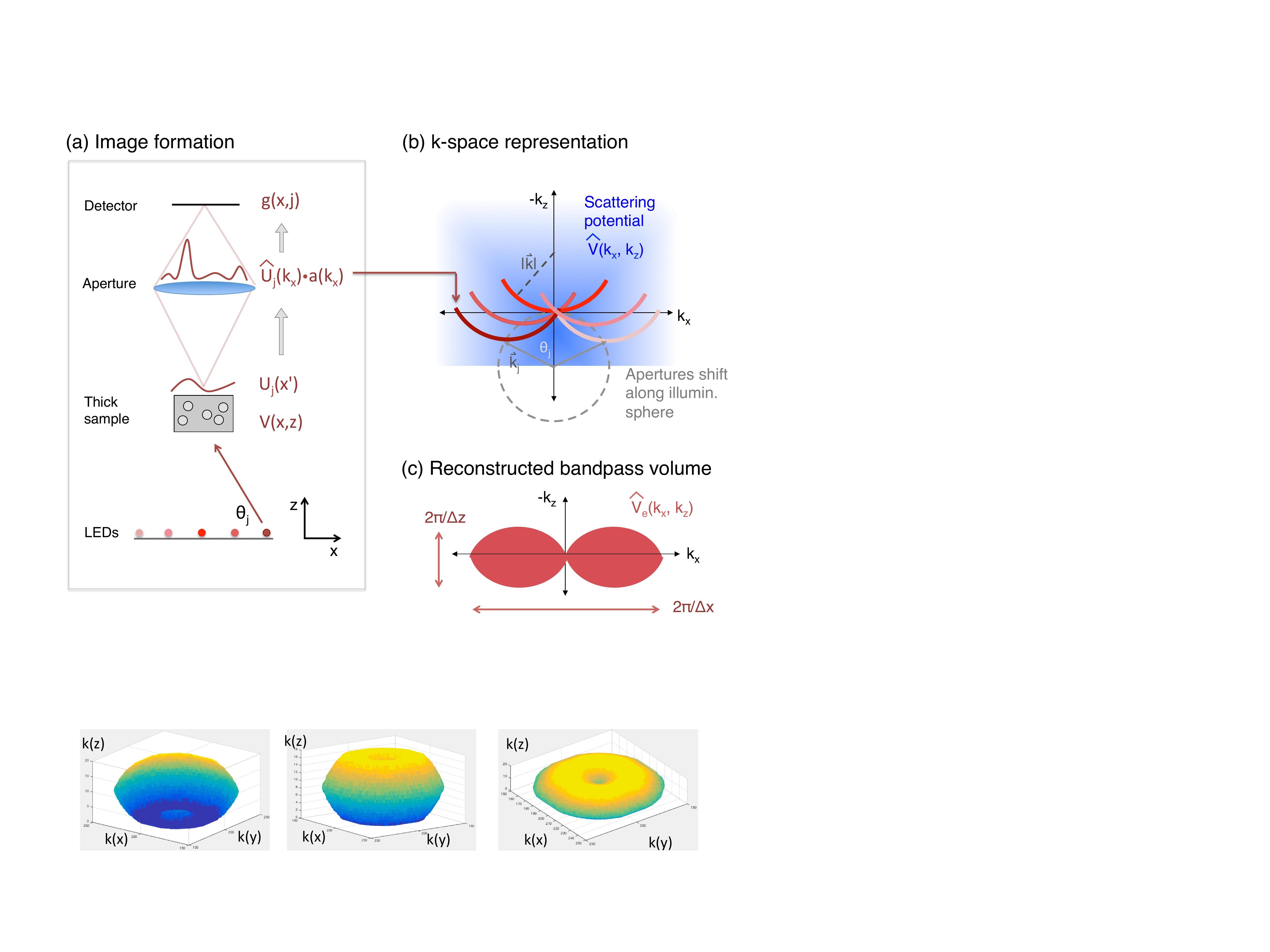}
\caption[2D]{Mathematical summary of FPT. (a) Diagram of the FPT setup in 2D. The field from the $j$th LED scatters through the sample and exits its top surface as $U_j(x')$. This field then propagates to form $\hat{U}_j(k_x)$ at the microscope back focal plane, where it is band-limited by the finite microscope aperture, $a(k_x)$. This band-limited field then propagates to the image plane, where its intensity is sampled to form the $j$th image. (b) Under the first Born approximation, each detected image is the squared magnitude of the Fourier transform of one colored "shell" in $(k_x,k_z)$ space. (c) By filling in this space with a ptychographic phase retrieval algorithm, FPT reconstructs the complex values within the finite bandpass volume $\hat{V}_e(k_x,k_z)$. The Fourier transform of this reconstruction yields our complex sample index of refraction map. }
\vspace{-0.2cm}
\label{geometry}
\end{figure} 

\section{Methods}
\label{sec:methods}

In this section, we develop a mathematical expression for our image measurements using the FPT framework, and then summarize our reconstruction algorithm. We describe our setup and reconstruction in 3D with the vector $\textbf{r}=(r_x,r_y,r_z)$ defining the sample coordinates and the vector $\textbf{k}=(k_x,k_y,k_z)$ defining the k-space (wavevector) coordinates (see Fig.~\ref{teaser}). 

\subsection{Image formation in FPT}

It is helpful to begin our discussion by introducing a quantity termed the scattering potential, which contains the complex index of refraction of an arbitrarily thick volumetric sample:
\begin{equation}
V(\textbf{r})=\frac{k}{4\pi}\left(n^{2}(\textbf{r})-n_b^2\right).
\label{potential}
\end{equation}
Here, $n(\textbf{r})$ is the spatially varying and complex refractive index profile of the sample, $n_b$ is the index of refraction of the background (which we assume is constant), and $k=2\pi/\lambda$ is the wavenumber in vaccuum. It is informative to point out that $n(\textbf{r})=n_r(\textbf{r})+in_i(\textbf{r})$, where $n_r$ is associated with the sample's refractive index and $n_i$ is associated with its absorptivity. We typically neglect the dependence of $n$ on $\lambda$ since we illuminate with quasi-monochromatic light. This dependence cannot be neglected when imaging with polychromatic light.

Next, to understand what happens to light when it passes through this volumetric sample, we define the complex field that results from illuminating the thick sample, $U(\textbf{r})$, as a sum of two fields: $U(\textbf{r})=U_i(\textbf{r})+U_s(\textbf{r})$. Here, $U_i(\textbf{r})$ is the field incident upon the sample (i.e., from one LED) and $U_s(\textbf{r})$ is the resulting field that scatters off of the sample. As detailed in \cite{Wolf69}, we may insert this decomposition into the scalar wave equation for light propagating through an inhomogeneous medium. Using Green's theorem, we may determine the scattered field as,
\begin{equation}
U_s(\textbf{r}') = \int G(|\textbf{r}'-\textbf{r}|)V(\textbf{r})U(\textbf{r}) d\textbf{r}.
\label{Greens1}
\end{equation}    
Here, $G(|\textbf{r}'-\textbf{r}|)$ is the Green's function connecting light scattered from various sample locations, denoted by $\textbf{r}$, to an arbitrary location $\textbf{r}'$. $V(\textbf{r})$ is the scattering potential from Eq.~\ref{potential}. Since $U(\textbf{r})$ is unknown at all sample locations, it is challenging to solve Eq.~\ref{Greens1}. Instead, it is helpful to apply the first Born approximation, which replaces $U(\textbf{r})$ in the integrand with $U_i(\textbf{r})$. This approximation assumes that $U_i(\textbf{r}) \gg U_s(\textbf{r})$. It is the first term in the Born expansion that describes the scattering response of an arbitrary sample~\cite{Wolf69}.

Our system sequentially illuminates the sample with an LED array, which contains $q=q_x \times q_y$ sources positioned a large distance $l$ from the sample (in a uniform grid, with inter-LED spacing $c$, see Fig.~\ref{teaser}). It is helpful to label each LED with a 2D counter variable $\left(j_x,j_y\right)$, where $-q_x/2\le j_x \le q_x/2$ and $-q_y/2\le j_y \le q_y/2$, as well as a single counter variable $j$, where $1\le j\le q$. Assuming each LED acts as a spatially coherent and quasi-monochromatic source (central wavelength $\lambda$) placed at a large distance from the sample, the incident field takes the form of a plane wave traveling at a variable angle such that $\theta_{jx}=\tan^{-1}(j_x \cdot c/l)$ and $\theta_{jy}=\tan^{-1}(j_y \cdot c/l)$ with respect to the $x$ and $y$ axes, respectively. We may express the $j$th field incident upon the sample as,
\begin{equation}
U_i^{(j)}(\textbf{r})=\textrm{exp}(i \textbf{k}_j \cdot \textbf{r}),
\label{incident}
\end{equation}  
where $\textbf{k}_j=\left(k_{jx},k_{jy},k_{jz}\right)=k \cdot \left(\sin{\theta_{jx}},\sin{\theta_{jy}},\sqrt{1-\sin^2{\theta_{jx}}-\sin^2{\theta_{jy}}} \right)$ is the wavevector of the $j$th LED plane wave. As $\theta_{jx}$ and $\theta_{jy}$ vary, $\textbf{k}_j$ will always assume values along a spherical shell in 3D $\left(k_x,k_y,k_z\right)$ space (i.e., the Ewald sphere), since the value of $k_{jz}$ is a deterministic function of $k_{jx}$ and $k_{jy}$.

After replacing $U(\textbf{r})$ in Eq.~\ref{Greens1} with $U_i^{(j)}(\textbf{r})$ in Eq.~\ref{incident}, and additionally approximating the Green's function $G$ as a far field response, the following relationship emerges between the scattering potential $V$ and the Fourier transform of the $j$th scattered field, $\hat{U}_s^{(j)}(\textbf{k})$, in the far field~\cite{Wolf69}:
 \begin{equation}
\hat{U}_s^{(j)}(\textbf{k}) = \hat{V}(\textbf{k} - \textbf{k}_j)
\label{Greens2}
\end{equation}
Here, $\hat{V}(\textbf{k})$, which we refer to as the k-space scattering potential, is the three-dimensional Fourier transform of $V(\textbf{r})$, $\textbf{k}$ denotes the scattered wavevector in the far field, and we have neglected constant multiplication factors for simplicity. The field scattered by the sample and viewed at a large distance, $\hat{U}_s^{(j)}(\textbf{k})$, is given by the values along a specific manifold (or spherical ``shell") of the k-space scattering potential, here written as $\hat{V}(\textbf{k} - \textbf{k}_j)$. We illustrate the geometric connection between $\hat{V}(\textbf{k}-\textbf{k}_j)$ and $\hat{U}_s^{(j)}(\textbf{k})$ for a 2D optical geometry in Fig.~\ref{geometry}(b). The center of the $j$th shell is defined by the incident wavevector, $\textbf{k}_j$. For a given shell center, each value of interest lies on a spherical surface at a radial distance set by $|\textbf{k}|=k$ (see colored arcs in Fig.~\ref{geometry}(b)). As $\textbf{k}_j$ varies with the changing LED illumination, the shell center shifts along a second shell with the same radius (since $\textbf{k}_j$ is itself constrained to lie on an Ewald sphere, see gray circle in Fig.~\ref{geometry}(b)). 

The goal of DT is to determine all complex values within the volume $\hat{V}$, from a set of $q$ scattered fields, $\{\hat{U}_s\}_{j=1}^{q}$. It is common to measure these scattered fields holographically~\cite{Lauer02,Sung09}. Each 2D holographic measurement maps to the complex values of $\hat{V}$ that lie along one 2D shell. Values from multiple measurements (i.e., the multiple shells in Fig.~\ref{geometry}(b)) can be combined to form a k-space scattering potential estimate, $\hat{V}_e$~\cite{Haeberle}. Nearly all stationary optical setups will yield only an estimate, since it is challenging to measure data from the entire k-space scattering potential without rotating the sample. Fig.~\ref{teaser}(c) and Fig.~\ref{geometry}(c) each display a typical measurable volume, also termed a bandpass, from a limited-angle illumination and detection setup. Once sampled, an inverse 3D Fourier transform of the band-limited $\hat{V}_e(\textbf{k})$ yields the desired complex scattering potential estimate, $V_e(\textbf{r})$, from which the quantitative index of refraction is directly obtained.

In FPT, we do not measure the scattered fields holographically. Instead, we use a standard microscope to detect image intensities and apply a ptychographic phase retrieval algorithm to solve for the unknown complex potential. The scattered fields in Eq.~\ref{Greens2} are defined at the microscope objective back focal plane (i.e., its Fourier plane), whose 2D coordinates $\textbf{k}_{2D}=(k_x,k_y)$ are conjugate to the microscope focal plane coordinates $(x,y)$. If we neglect the effect of the constant background plane wave term (i.e., $U_i$ in the sum $U=U_i+U_s$), we may now write the $j$th shifted field at our microscope back focal plane as, $\hat{U}^{(j)}(\textbf{k}_{2D})=\hat{V}\left(\textbf{k}_{2D} - \textbf{k}_{j2D}, k_z-k_{jz}\right)$. These new coordinates highlight the 3D to 2D mapping from $\hat{V}$ to $\hat{U}$, where again $k_z=\sqrt{k-k_x^2-k_y^2}$ is a deterministic function of $\textbf{k}_{2D}$, and the same applies between $k_{jz}$ and $\textbf{k}_{j2D}$.

Each shifted scattered field is then bandlimited by the microscope aperture function, $a(\textbf{k}_{2D})$, before propagating to the image plane. The limited extent of $a(\textbf{k}_{2D})$ (defined by the imaging system NA) sets the maximum extent of each shell along $k_x$ and $k_y$. The $j$th intensity image acquired by the detector is given by the squared Fourier transform of the bandlimited field at the microscope back focal plane: 
\begin{equation}
g(x,y,j)=\left| F\left[\hat{V}\left(\textbf{k}_{2D} - \textbf{k}_{j2D}, k_z-k_{jz}\right) \cdot a(\textbf{k}_{2D}) \right] \right|^2.
\label{datamatrix2}
\end{equation}
Here, $F$ denotes a 2D Fourier transform with respect to $\textbf{k}_{2D}$ and we neglect the effects of magnification (for simplicity) by assuming the image plane coordinates match the sample plane coordinates, $(x,y)$. The goal of FPT is to determine the complex 3D function $\hat{V}$ from the real, positive data matrix, $g(x,y,j)$. A final 3D Fourier transform of $\hat{V}$ yields the desired scattering potential, and subsequently the refractive index distribution, of the thick sample.

\subsection{FPT reconstruction algorithm}

Eq.~\ref{datamatrix2} closely resembles the data matrix measured by Fourier ptychography (FP, see~\cite{Zheng13}). Now, however, intensities are sampled from a volumetric function along shells in a 3D space (i.e., the curves in Fig.~\ref{geometry}). We use an iterative reconstruction procedure, mirroring that from FP~\cite{Zheng13}, to ``fill in" the k-space scattering potential with data from each recorded intensity image. Just like FP requires a certain amount of data redundancy (i.e., overlap in k-space) to accurately recover the unknown optical phase, FPT also requires overlap between shell regions in 3D k-space. Since our discretized k-space now has an extra dimension, overlap is less frequent and more images are required for successful algorithm convergence. A coarser k-space discretization, a smaller LED array pitch and/or a larger array-sample distance along $z$ will help increase overlap. As we demonstrate experimentally, several hundred images are sufficient for a complex reconstruction that contains approximately 30 unique axial slices. 

It is important to select the correct limits and discretization of 3D k-space (i.e., the FOV and resolution of the complex sample reconstruction). The maximum resolvable wavevector along $k_x$ and $k_y$ is proportional to $k(\textrm{NA}_o + \textrm{NA}_i)$, where $\textrm{NA}_o$ is the objective NA and $\textrm{NA}_i$ is maximum NA of LED illumination. This lateral spatial resolution limit matches FP~\cite{Ou15a}. The maximum resolvable wavevector range along $k_z$ is also determined as a function of the objective and illumination NA as, $k_z^{\textrm{max}}=k\left( 2 - \sqrt{1-\textrm{NA}_o^2} - \sqrt{1-\textrm{NA}_i^2} \right)$. This relationship is easily derived from the geometry of the k-space bandpass volume in Fig.~\ref{geometry}, as shown in \cite{Lauer02}. We typically specify the maximum imaging range along the axial dimension, $z_{\textrm{max}}$, to equal approximately twice the expected sample thickness. This then sets the discretization level along $k_z$: $\Delta k_z=2\pi/z_{\textrm{max}}$. The total number of resolved slices along $z$ is set by the ratio $k_z^{\textrm{max}}/\Delta k_z$.  

We now summarize the FPT reconstruction algorithm in the following 5 steps:

\begin{enumerate}
\item Initialize a discrete estimate of the unknown k-space scattering potential, $\hat{V}_e(\textbf{k})$, selecting the appropriate 3D array size following the discussion above. Either a single raw image may be padded along all three dimensions and then Fourier transformed for this initialization, or the raw intensity image set may be used to form a refocused light field~\cite{Tian15}. A constant matrix is also often an adequate initialization. 

\item For $j=1$ to $q$ images, compute the center coordinate, $\textbf{k}_j$, and select its associated shell (radius $k$, maximum width $2k \cdot \textrm{NA}_o$). This selection process samples a discrete 2D function, $\hat{d}_j(k_x,k_y)$, from the 3D k-space volume. The selected voxels must partially overlap with voxels from adjacent shells. Currently, no interpolation is used to map voxels from the discrete shell to pixels within $\hat{d}_j(k_x,k_y)$.     

\item Fourier transform $\hat{d}_j(k_x,k_y)$ to the image plane to create $d_j(x,y)$, and constrain its amplitudes to match the measured amplitudes from the $j$th image. For example, the update may take the simple form, $d_j'(x,y)=\sqrt{g(x,y,j)} \cdot d_j(x,y)/|d_j(x,y)|$. More advanced alternating projection-based updates are also available~\cite{Marchesini07}.

\item Inverse 2D Fourier transform the image plane update, $d_j'(x,y)$, back to 2D k-space to form $\hat{d}_j'(k_x,k_y)$. Use the values of $\hat{d}_j'(k_x,k_y)$ to replace the voxel values of $\hat{V}_e(\textbf{k})$ at locations where voxel values were extracted in step 2.   

\item Repeat steps 2-4 for all $j=1$ to $q$ images. This completes one iteration of the FPT algorithm. Continue for a fixed number of iterations, or until satisfying some error metric. At the end, 3D inverse Fourier transform $\hat{V}_e(\textbf{k})$ to recover the complex scattering potential, $V_e(\textbf{r})$.

\end{enumerate} 

In practice, we also implement a pupil function recovery procedure~\cite{Ou14e} as we update each extracted shell from k-space. This allows us to simultaneously estimate and remove possible aberrations present in the microscope back focal plane.

\begin{figure}[]
\centering
\includegraphics[width=.99\columnwidth]{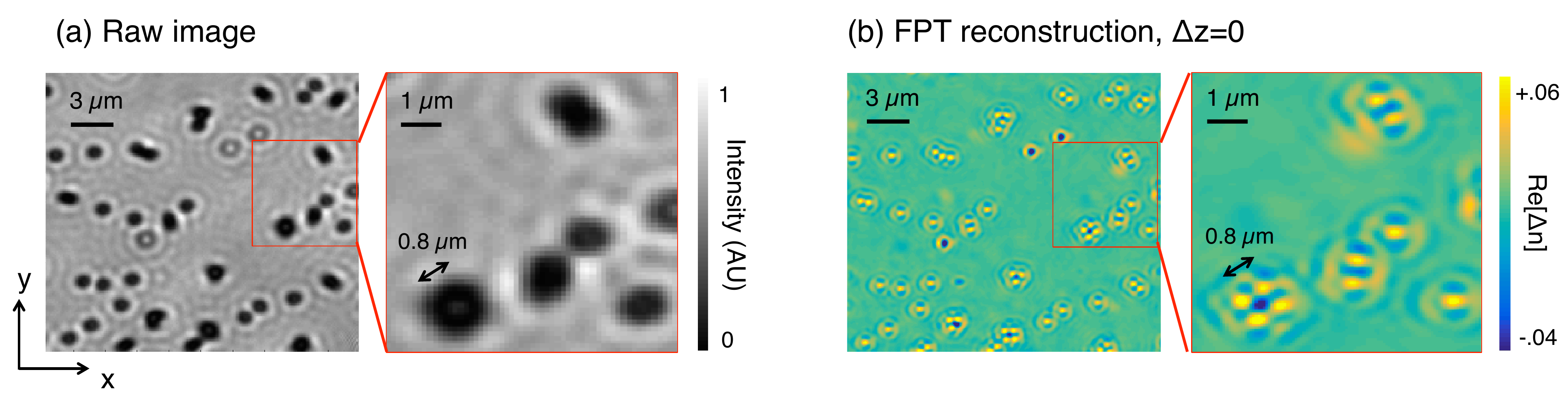}
\caption[2D]{FPT improves the lateral resolution of a standard microscope. (a) A single raw image of a layer of 0.8 $\mu$m microspheres immersed in oil, where beads within each cluster are not resolved. (b) The real component of the index of refraction from one slice (out of 30) for our FPT reconstruction ($\Delta z=0$ slice), which clearly resolves each microsphere. }
\vspace{-0.2cm}
\label{ms_res}
\end{figure}

\section{Results}
\label{sec:results}

We experimentally verify our reconstruction technique using a standard microscope outfitted with an LED array. The microscope uses an infinity corrected objective lens ($\textrm{NA}_o=0.4$, Olympus MPLN, 20X), to image onto a digital detector containing 4.54 $\mu$m pixels (Prosilica GX 1920, 1936$\times$1456 pixel count). The LED array contains 31 $\times$ 31 surface-mounted elements (model SMD3528, center wavelength $\lambda=$632 nm, 4 mm LED pitch, 150 $\mu$m active area diameter). For this first demonstration, we position the LED array approximately 135 mm beneath the sample to create a maximum illumination NA of $\textrm{NA}_i=$0.41. This leads to an effective lateral NA of $\textrm{NA}_o+\textrm{NA}_i=0.81$, and a lateral resolution gain along $(x,y)$ of slightly over a factor of 2 (from a 1.6 $\mu$m minimum resolved spatial period in the raw images to a 0.78 $\mu$m minimum resolved spatial period in the reconstruction). The associated axial resolution is computed at 3.7 $\mu$m, and we reconstruct quantitative sample information across a total depth range of approximately $z_{\textrm{max}}=$110 $\mu$m (approximately 20 times larger than the stated objective lens DOF of 5.8 $\mu$m). 

For most of the reconstructions presented below, we capture and process $q=675$ images from the same fixed pattern of LEDs. We typically use the following parameters for FPT reconstruction: each raw image is cropped to 1000 $\times$ 1000 pixels, the reconstruction voxel size is set at 0.39 $\times$ 0.39 $\times$ 3.7 $\mu$m$^{3}$ for sampling at the Nyquist-Shannon rate, the reconstruction array contains approximately 2100 $\times$ 2100 $\times$ 30 voxels, and the algorithm runs for 5 iterations.

\begin{figure}[]
\centering
\includegraphics[width=.8\columnwidth]{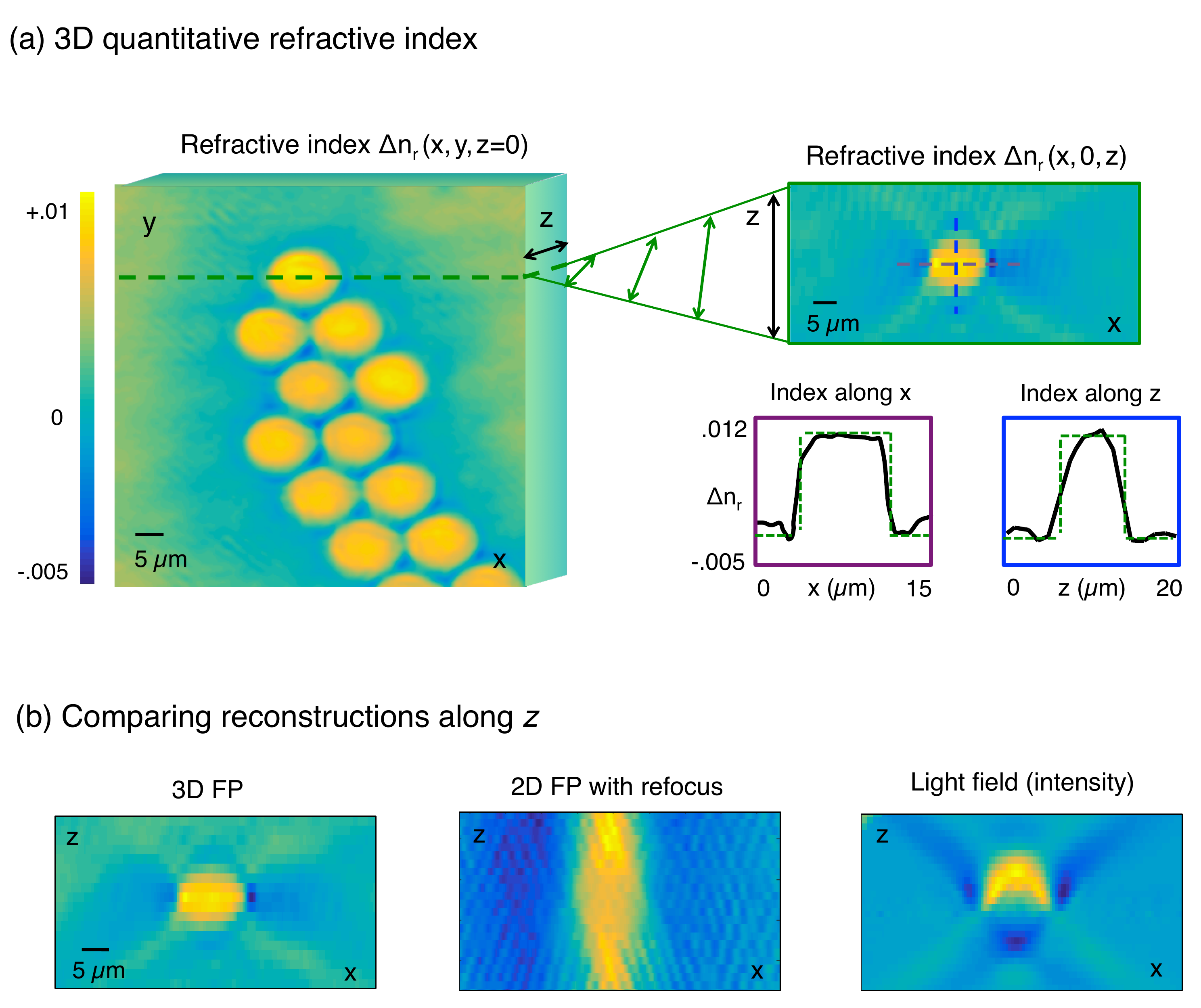}
\caption[2D]{FPT quantitatively measures the complex index of refraction of samples in 3D. (a) Tomographic reconstruction of 12 $\mu$m microspheres immersed in oil, where we show a lateral ($\Delta z=0$) slice on the left, an axial ($\Delta y=25$ $\mu$m) slice on the right, and 1D plots of the index shift along both $x$ and $z$, demonstrating quantitative performance. (b) We use the same dataset to obtain an FP reconstruction and propagate the result along $z$ (middle), and also perform light field refocusing (right). Our FPT reconstruction (left) offers the closest match to the expected axial profile of a spherical bead. }
\vspace{-0.2cm}
\label{ms_quant}
\end{figure} 

\subsection{Quantitative verification}

We include three different quantitative verifications of FPT performance using polystyrene microspheres as reference targets. First, we verify the ability of FPT to improve lateral image resolution. This matches the goal of FP for thin 2D samples. The sample consists of 800 nm diameter microspheres (index of refraction $n_s=1.59$) immersed in oil (index of refraction $n_o=1.515$). We highlight a small group of these microspheres in Fig.~\ref{ms_res}. The single raw image in (a) (generated from the center LED) cannot resolve the individual spheres gathered in small clusters. Based upon the coherent Sparrow limit for resolving two points ($0.68\lambda/\textrm{NA}_{o}$) this raw image cannot resolve points that are closer than 1.1 $\mu$m. After FPT reconstruction, we obtain the complex index of refraction in Fig.~\ref{ms_res}(b), where here we show the real component of the recovered index. The FPT reconstruction along the $\Delta z=0$ slice clearly resolves the spheres within each cluster. This 800 nm distance is close to the expected Sparrow limit for FPT of $0.68\lambda/\left(\textrm{NA}_{o}+\textrm{NA}_{i}\right)=540$ nm. The ringing features around each sphere indicate a sinc-like point-spread function, as we expect theoretically, and this ringing constructively interferes to form the undesired dip feature at the center of each cluster.   

Second, we check the quantitative accuracy of FPT by imaging microspheres that extend across more than just a few reconstruction voxels. Fig.~\ref{ms_quant} displays a reconstruction of 12 $\mu$m diameter microspheres (index of refraction $n_s=1.59$) immersed in oil (index of refraction $n_o=1.58$). We use the same data capture and post-processing steps as in Fig.~\ref{ms_res}, and display a cropped section (200$\times$200$\times$15 voxels) of the full reconstruction. We again display the real (non-absorptive) component of the recovered index across both a lateral slice (along the $\Delta z=0$ plane) and a vertical slice (along the $\Delta y=25$ $\mu$m plane). We also include detailed 1D traces along the center of the vertical slice. 

Three observations are noteworthy regarding this experiment. First, the measured index shift approximately matches the expected shift of $\Delta n=n_s-n_o=0.01$ across the entire bead, thus demonstrating quantitatively accurate performance. Second, for each 1D trace through the center of each microsphere, we would ideally expect a perfect rect function (from $\Delta n=0$ to $\Delta n=0.01$ and then back down). This is unlike 2D FP, which reconstructs the phase delay though each sphere, leading to a parabolic function (due to their varying thickness along the optical axis). While the system can resolve an approximate step function through the center of the sphere along the lateral ($x$) dimension, it is not a step function function along the axial ($z$) direction. This is caused by the limited extent of the measurable volume of 3D k-space (i.e., the limited bandpass). The ``missing cone" of information, primarily surrounding the $k_z$ axis, creates a noticeably wide point-spread function along $z$, which leads to its distinct sinc shape. Various methods are available to computationally fill in the missing cone using prior sample information~\cite{Tam81,Medoff83}. 

For our third observation, we compare FPT with two alternative techniques for 3D imaging in Fig.~\ref{ms_quant}(c). First, we use the same dataset to perform 2D FP, and then attempt to holographically refocus its complex 2D solution. We obtain this solution using the same number of images ($q=675$) and with the FP procedure in \cite{Zheng13}, after focusing the objective lens at the axial center of the 12 $\mu$m microspheres. The ``out of focus noise" above and below the plane of the microsphere, created by digital propagation of the complex field via the angular spectrum method, quite noticeably hides its spherical shape. Second, we interpret the same raw image set as a light field and perform light field refocusing~\cite{Levoy09}. While the refocused light field approximately resolves the outline of microsphere along $z$, it does not offer a quantitative picture of the sample interior, nor a measure of its complex index of refraction. The areas above the microsphere are very bright due to its lensing effect (i.e., the light field displays the optical intensity at each plane, and thus displays high energy where the microsphere focuses light). FPT thus appears more accurate here than these two alternatives.

\begin{figure}[]
\centering
\includegraphics[width=.75\columnwidth]{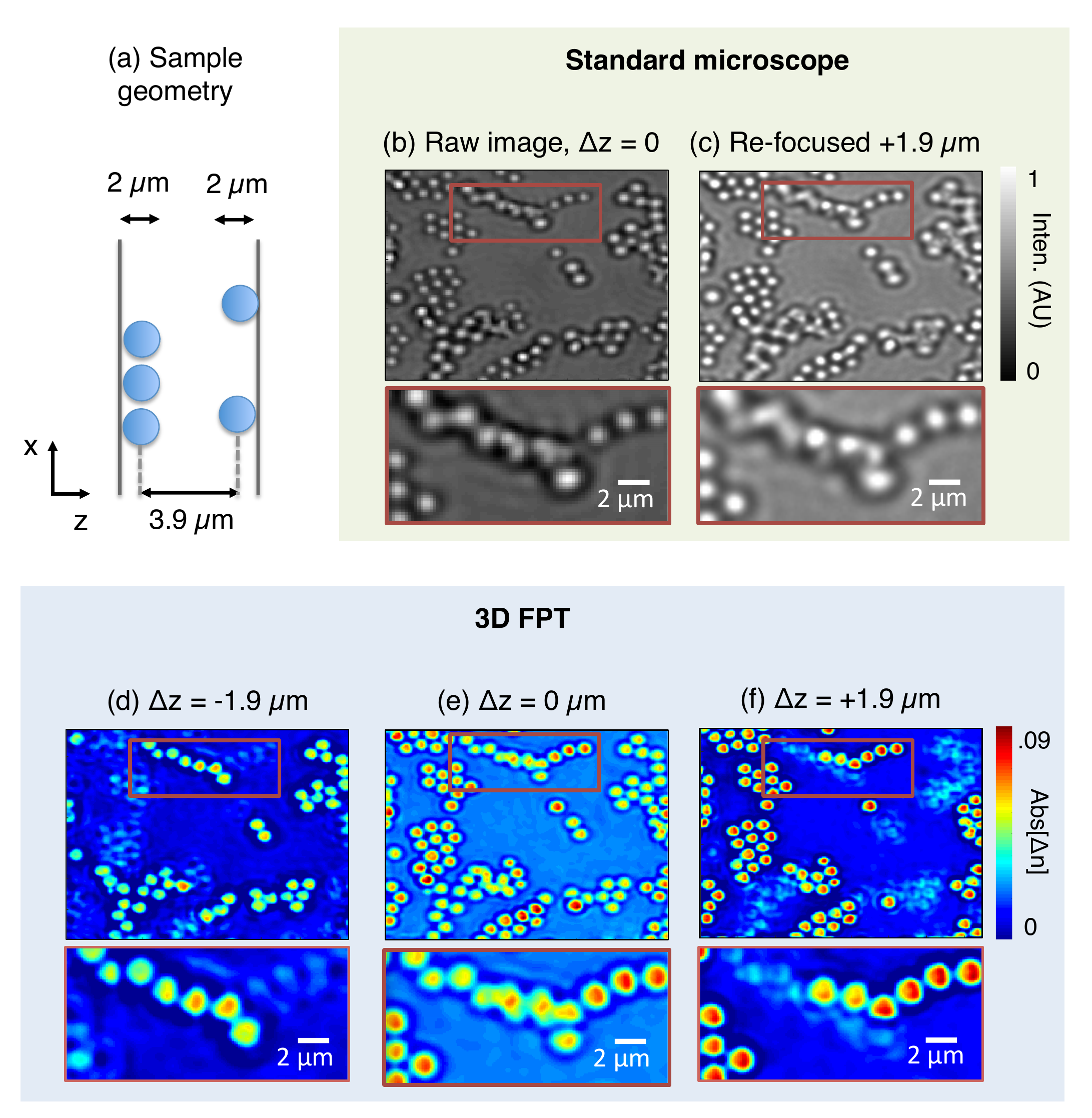}
\caption[2D]{Experimentally measuring the axial resolution of FPT. (a) The reconstructed sample contains two layers of microspheres separated by a thin layer of oil. Raw images (b) focused at the center of the two layers and (c) on the top layer do not clearly resolve overlapping microspheres (e.g., in red box). (d)-(e) Slices of the FPT tomographic reconstruction, showing $|\Delta n|$, clearly resolve each sphere within the two individual sphere layers.  }
\vspace{-0.2cm}
\label{ms_sep}
\end{figure} 

For our third and final quantitative test, we verify the axial resolution of FPT along $z$. Here, we prepare a sample containing two closely separated layers. Each layer contains 2 $\mu$m microspheres ($n_s=1.59$) distributed across the surface of a glass slide, which we sandwich together with oil in between ($n_o=1.515$). The separation between the two microsphere layers, measured from the center of each sphere along $z$, is 3.9 $\mu$m (i.e., the separation between the microscope slide surfaces is 5.9 $\mu$m, as diagrammed in Fig.~\ref{ms_sep}(a)). The 3.9 $\mu$m center-to-center distance is close to the expected axial resolution limit of 3.7 $\mu$m for the FPT microscope. 

Conventional microscope images of the sample, using the center LED for illumination, are in Fig.~\ref{ms_sep}(b)-(c), where we focus to the center of the two layers ($\Delta z=0$) as well as the top microsphere layer ($\Delta z=1.9$ $\mu$m) in an attempt to distinguish the two separate layers. At the top of each image (where microspheres in the two layers overlap), it is especially hard to resolve each sphere or determine which sphere is in a particular layer. These challenges are due in part to the limited ability to measure just the optical intensity at each plane, instead of the complex refractive index of the sample. 

Next, we return the focus to the $\Delta z=0$ plane and implement FPT. We display three slices of our 3D scattering potential reconstruction in Fig.~\ref{ms_sep}(d)-(f). Here, we show the absolute value of the potential near the plane of the top layer, at the center, and near the plane of the bottom layer. The originally indistinguishable spheres within the top and bottom layers are now clearly resolved in each $z$-plane. Due to the system's limited axial resolution, the reconstruction at the middle plane ($\Delta z=0$) still shows the presence of spheres from both layers. Comparing Fig.~\ref{ms_sep}(b)-(c) with Fig.~\ref{ms_sep}(e)-(f) clearly maintains that the axial resolution of FPT is sharper than manual refocusing. Not only is each layer clearly distinguishable (as predicted theoretically), but we now also have quantitative information about the sample's complex refractive index.

\begin{figure}[]
\centering
\includegraphics[width=.8\columnwidth]{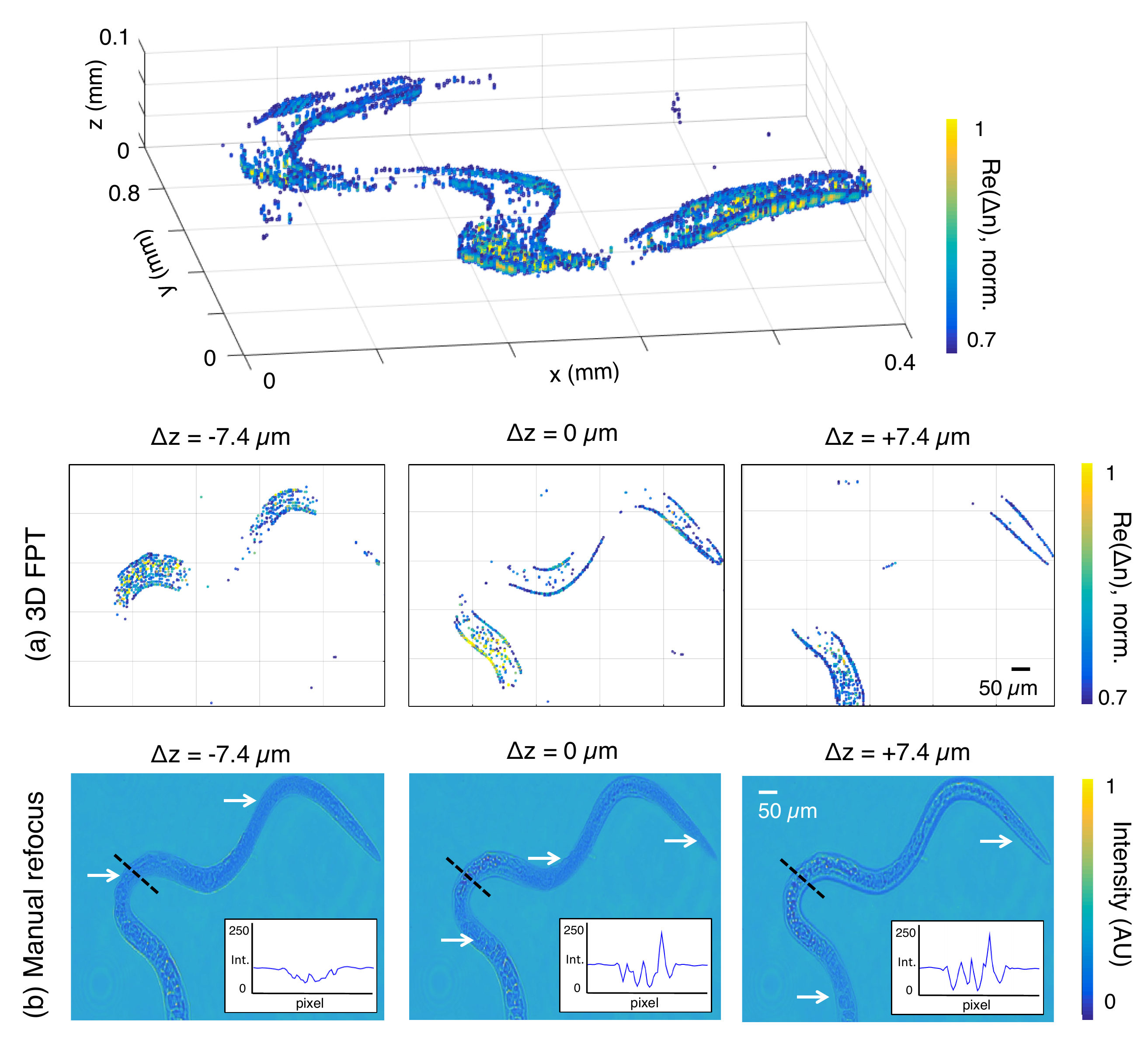}
\caption[2D]{Tomographic reconstruction of a \textit{trichinella spiralis} parasite. (a) The worm's curved trajectory is clearly resolved within the various $z$-planes. (b) Refocusing the same distance to each respective plane does not clearly distinguish each in-focus worm segment (marked by white arrows). Since the worm is primarily transparent, in-focus worm sections exhibit minimal intensity contrast, presenting significant challenges for segmentation (see inset plots of intensity along each black dash, where the section is in-focus in the image on left). FPT, on the other hand, exhibits maximum contrast at each voxel containing the worm.}
\vspace{-0.2cm}
\label{worm}
\end{figure} 

\subsection{Biological experiments}

Next, we use FPT to reconstruct the 3D complex refractive index of two different thick biological specimens. Since the exact composition of these specimens is unknown, it is challenging to verify the quantitative accuracy of our reconstructions here, especially given the accuracy of first Born approximation is only guaranteed up to a total phase shift of approximately one wavelength. However, we will point out the multiple qualitative benefits of FPT in these examples. 

Our first tomographic reconstruction is of a \textit{trichinella spiralis} parasite (Fig.~\ref{worm}). Here, since the worm extends along a larger distance than the width of our detector, we performed FPT twice, shifting the FOV between to capture the left and right side of the worm with 10\% overlap between. We then merged each tomographic reconstruction together with a simple averaging operation (matching that from FP~\cite{Zheng13}). The total captured volume here is 0.8 mm$\times$0.4 mm$\times$110 $\mu$m. If we were to replace our current digital detector with one that occupied the entire microscope FOV, we would increase our fixed imaging volume to 2.2 mm$\times$2.2 mm$\times$110 $\mu$m, and obtain tomograms that each contain approximately $10^9$ voxels per acquisition.

We display a thresholded 3D scattering potential reconstruction of the parasite at the top of Fig.~\ref{worm} (real component, threshold applied at $\textrm{Re}[\Delta n]> 0.7$ after $|\Delta n|$ normalized to 1, under-sampled for clarity). Its 3D curved trajectory is especially clear in the three separate $z$-slices of the reconstructed tomogram in Fig.~\ref{worm}(a). The two downward bends in the parasite body are lower than the upward bend in the middle, as well as at its front and back ends. It is very challenging to resolve these depth-dependent sample features by simply refocusing a standard microscope. Fig.~\ref{worm}(b) displays such an attempt, where the same three $z$ planes are brought into focus manually. Since the sample is primarily transparent, in-focus areas in each standard image actually exhibit minimal contrast, as marked by arrows in Fig.~\ref{worm}(b). We plot the intensity through one worm section (black dash) in three insets. The intensity contrast drops by over a factor of 2 at in-focus locations, which will prevent the success of depth segmentation techniques (e.g., focal stack deconvolution~\cite{Agard84}). Since FPT effectively offers phase contrast, points along the parasite within its reconstruction voxels instead show maximum contrast, which enables direct segmentation via thresholding, as achieved in the top plot.

\begin{figure}[]
\centering
\includegraphics[width=.75\columnwidth]{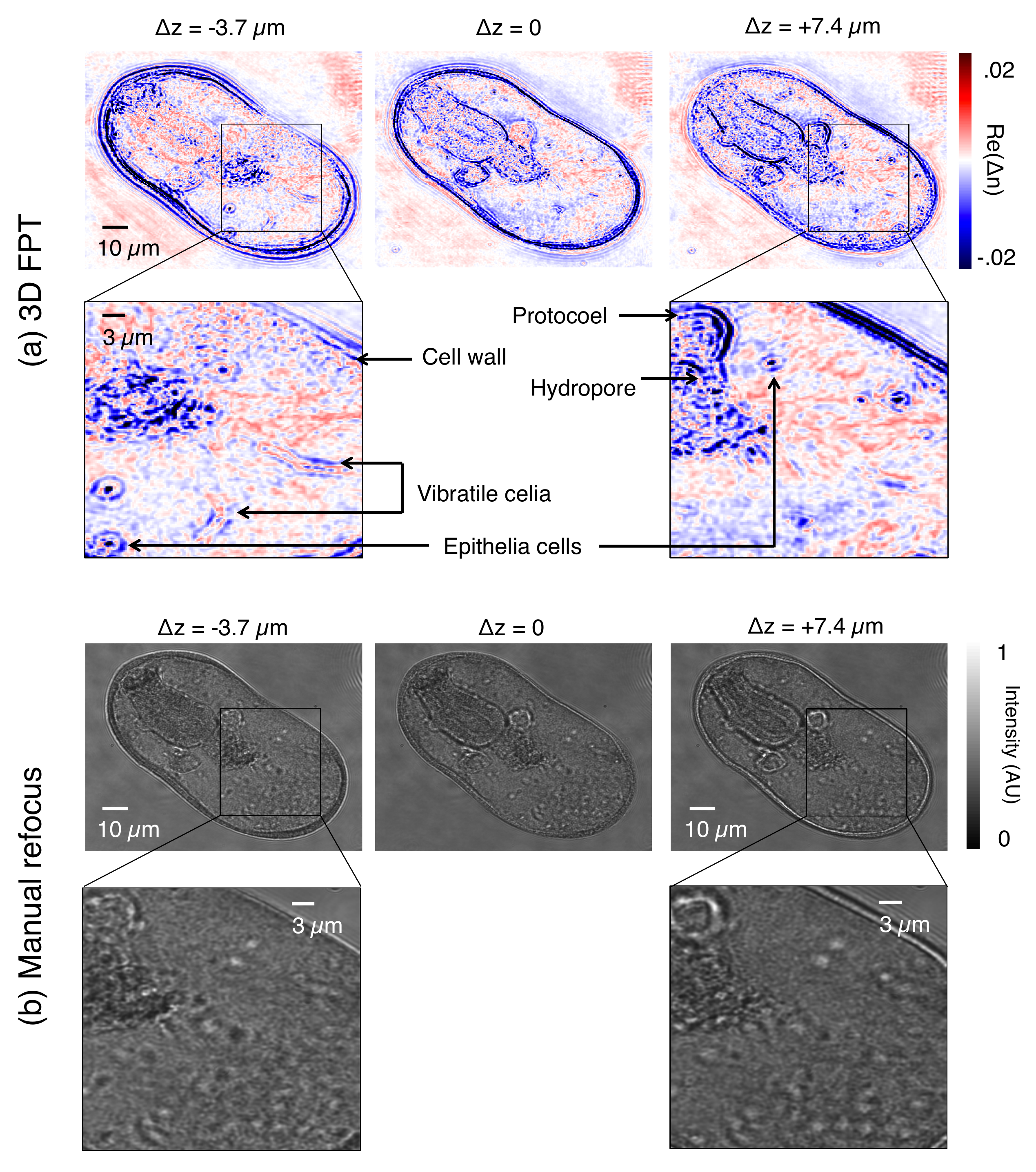}
\caption[2D]{3D reconstruction of a starfish embryo at the larval stage. (a) Three different axial planes of the FPT tomogram show significantly different features within the larva. For example, the protocoel is completely missing from the $\Delta z=-3.7$ $\mu$m plane. Likewise, what we expect to be the developing vibratile celia (see lower right) are \textit{only} visible in the $\Delta z=-3.7$ $\mu$m plane. (b) This type of axial information, and even certain structures (e.g., the vibratile celia and various epithelia cells, marked in (a)) are completely missing from standard microscope images after manually refocusing to each plane of interest.}
\vspace{-0.2cm}
\label{starfish}
\end{figure} 

For our second 3D biological example, we tomographically reconstruct a starfish embryo at its larval stage (Fig.~\ref{starfish}). Here, we again show three different closely spaced $z$-slices of the reconstructed scattering potential ($\textrm{Re}[\Delta n]$, no thresholding applied). Each $z$-slice contains sample features that are not present in the adjacent $z$-slices. For example, the large oval structure in the upper left of the $\Delta z=0$ plane, which is a developing stomach, nearly completely disappears in the $\Delta z=-3.3$ $\mu$m plane. Now at this $z$-slice, however, small structures, which we expect to be the developing vibratile celia as well as epithelia cells at various locations~\cite{Agassiz}, clearly appear in the lower right. Both the particular plane of the developing stomach and even the presence of the vibratile celia are completely missing from the refocused images. This is due to the inability of the standard microscope to segment each particular plane of interest, the inability to accurately reconstruct transparent structures without a phase contrast mechanism, and an inferior lateral $(x,y)$ resolution with respect to FPT.  

\section{Conclusion}
\label{sec:conclusion}

The FPT method performs diffraction tomography using intensity measurements, captured with a standard microscope and an LED illuminator. Its reconstruction algorithm extends previous work with FP to now operate in 3D. The current system offers a lateral resolution of approximately 400 nm at the Nyquist-Shannon sampling limit (550 nm at the Sparrow limit and 800 nm full period limit), and an axial resolution of 3.7 $\mu$m at the sampling limit. The maximum axial extent attempted thus far was 110 $\mu$m along $z$, which leads to approximately one giga-voxel of complex sampling points per acquisition if imaging over the total microscope FOV (2.2 $\times$ 2.2 mm). We demonstrated quantitative measurement of the complex index of refraction within thick biological specimens. 
 
We believe that FPT can be significantly improved with additional experimental development. First, a better LED array geometry will enable a higher angle of illumination to improve resolution. Second, we set the number of captured images here to match previously determined data redundancy requirements~\cite{Bunk08}. However, we have observed that reconstructions are successful with much fewer images than otherwise expected. Along with using a multiplexed illumination strategy~\cite{Tian14a}, this may help significantly speed up tomogram capture time. Third, we set our reconstruction range along the $z$-axis somewhat arbitrarily at 110 $\mu$m. We expect the ability to further extend this axial range in future setups. 

Alternative computational approaches may also improve FPT. Here, we list a number of possible directions. First, we adopted the well-known alternating projections method (i.e., the ePIE algorithm~\cite{Maiden09}) for ptychographic update. Other methods, such as convex-based approaches~\cite{Horstmeyer15c}, can perform better in the presence of noise. Second, alternative approximations are also available to solve the first Born approximation~\cite{Devaney81}. Third, a big detriment to resolution is currently the missing cone in 3D k-space, and various methods are available to fill this cone in, e.g., by assuming the sample is positive-only, sparse, or of a finite spatial support~\cite{Tam81,Medoff83}. Finally, there are already suggested methods to solve for the full Born series, which take into account the effects of multiple scattering~\cite{Wang89}. Connections between this type of multiple scattering solver, recent methods applying the multi-slice approximation~\cite{Tian15,Li15,Kamilov15}, and FPT may lead to successful reconstruction of increasingly turbid samples.

\section*{Funding Information}
National Institute of Health (NIH) (1DP2OD007307-01); The Caltech Innovation Initiative (CI2) internal grant program (13520135).

\section*{Acknowledgments}
The authors thank X. Ou, J. Chung, J. Brake and G. Zheng for helpful discussions and technical support.

\section*{Supplemental Documents}
See Supplemental Movies for full tomographic reconstructions.

\end{document}